\documentclass[preprint,aps,showpacs,11pt]{revtex4-1}
\usepackage{hyperref}
\usepackage{amsmath}
\usepackage{amsfonts}
\usepackage{amssymb}
\usepackage[dvipdf]{graphicx}
\def\gsim{\lower.7ex\hbox{$\;\stackrel{\textstyle>}{\sim}\;$}}
\def\lsim{\lower.7ex\hbox{$\;\stackrel{\textstyle<}{\sim}\;$}}
\usepackage{graphicx}
\usepackage{dcolumn}
\usepackage{bm}
\usepackage{xcolor}
\usepackage{hyperref}
\usepackage[mathlines]{lineno}
\def\gsim{\lower.7ex\hbox{$\;\stackrel{\textstyle>}{\sim}\;$}}
\def\lsim{\lower.7ex\hbox{$\;\stackrel{\textstyle<}{\sim}\;$}}
\usepackage{color}
\usepackage[utf8]{inputenc}
\usepackage{soul}
\usepackage{ulem}
\definecolor{red}{rgb}{1,0,0}
\usepackage{epsfig,bbm}
\newcommand{\ba}{\begin{array}}
\newcommand{\ea}{\end{array}}
\newcommand{\be}{\begin{equation}}
\newcommand{\ee}{\end{equation}}
\newcommand{\bea}{\begin{eqnarray}}
\newcommand{\eea}{\end{eqnarray}}

\newcommand{\M}{\mathcal{M}}

\begin{document}
\title{$S(3)$ flavoured Higgs model trilinear self-couplings}
\author{E. Barradas-Guevara}
 \email{barradas@fcfm.buap.mx}
\affiliation{Fac. de Cs. F\'{\i}sico Matem\'aticas, Benem\'erita Universidad Aut\'onoma de Puebla, Apdo. Postal 1152, Puebla, Pue.  72000, M\'exico.}%
\author{O. F\'elix-Beltr\'an}
\affiliation{Fac. de Cs. de la Electr\'onica, Benem\'erita Universidad Aut\'onoma de Puebla, Apdo. Postal 542, Puebla, Pue. 72570, M\'exico.}
 \altaffiliation[Also at ]{olga\_flix@ece.buap.mx}
\author{E. Rodr\'{\i}guez-J\'auregui}
 \email{ezequiel.rodriguez@correo.fisica.uson.mx}
\affiliation{Departamento de F\'{\i}sica, Universidad de Sonora, Apdo. Postal 1626, Hermosillo, Son.  83000, M\'exico.}%

\date{\today}

\begin{abstract}
In this work, we analyze  the Higgs sector of the minimal $S(3)$-invariant extension of the Standard
Model is performed. Considering three Higgs fields, which are SU(2) doublets, and CP invariant, we compute the exact and analytical physical Higgs boson masses in terms of the Higgs potential parameters and the scalar Higgs matrix rotation angle $\theta_S$ and $w_3$ ($\tan\theta_P=\tan\theta_C=\tan^{-1}\omega_3$), related to  the pseudoscalar and charged Higgs matrix rotation angles $\theta_P$ and $\theta_C$ respectively. Furthermore, within this model we can also write down in an explicit form the trilinear self-couplings $\lambda_{ijk}$ in terms of the Higgs masses and two free parameters, $\theta_S$ and $w_3$.  Moreover, we show that the Higgs masses and  trilinear Higgs bosons self-couplings are closely linked to the  Higgs potential structure given by the discrete symmetry $S(3)$, which can be  helpful to distinguish this model from other extensions. In our analysis the lightest Higgs boson mass is taken to be fixed to 125 GeV.  In concordance with the results reported in the literature for other Standard Model extensions, it find that the numerical values $\lambda_{ijk}$ of the minimal $S(3)$-invariant extension of the Standard Model  are significantly different from the trilinear Higgs self-coupling of the Standard Model.
\end{abstract}


\maketitle

%\tableofcontents

%%%%%%%%%%%%%%%%%%%
\section{Introduction}\label{sec:int}
%%%%%%%%%%%%%%%%%%%
In the Standard Model (SM) the electroweak gauge bosons and the fundamental matter particles acquire masses through the interaction with a scalar field. The Higgs boson is an essential part of the SM upon the electroweak spontaneous symmetry  breaking (SSB).  The Higgs mechanism~\cite{Higgs:1964pj} of SSB in the context of the SM is being tested for internal consistency.
The discovery of the Higgs boson candidate on July 2012  is one of the major goals of the  Large Hadron Collider (LHC)~\cite{Aad:2012tfa,Chatrchyan:2013lba,Chatrchyan:2012ufa}.  Although the mass of the Higgs boson was measured with a good approximation value (125 GeV), several properties as the spin, the Higgs self-couplings and fermionic-bosonic Higgs boson couplings must be measured.  Several works on this were done by  A. Djouadi {\it et al.}~\cite{Djouadi:1999rca,Djouadi:1999gv}, J. Baglio {\it et al.}~\cite{Baglio:2012np}, P. Osland {\it et al.}~\cite{Osland:1998hv,Osland:1999qw,Osland:1999ae,Osland:2008aw}, and as well as \cite{Tian:2013yda,Okada:2013rha}.

 These days it remains  to be seen if the new boson with a mass of 125 GeV, observed in the CMS experiment at LHC  and corresponding to the scalar in the SM, is of ultimate importance to analyze the mechanism of spontaneous symmetry breaking in models which include a larger number of scalar bosons~\cite{Djouadi:1999rca,Chatrchyan:2012ufa}.
Different aspects of models including three and more Higgs doublets have also been studied, with and without discrete symmetries 
(see for instance~\cite{Lendvai:1981wn}). Several extensions of the SM, such as the $S(3)$ extended Standard Model~\cite{Teshima:2005bk,Kobayashi:2003fh,Kubo:2003pd,Kubo:2003iw}, the Minimal Supersymmetric Standard Model (MSSM)~\cite{Haber:1984rc}, and the general two Higgs doublet model (2HDM)~\cite{Branco:2011iw,Crivellin:2013wna} have a more complicated Higgs structure.
 An  extended Standard Model with three $SU(2)$  Higgs doublets  has a more complicated Higgs potential structure with 54 free parameters.

On the other hand, in the absence of mass the Lagrangian is chiral and invariant with respect to any permutation of the left and right quark and lepton fields. As a consequence, the left and right quark and lepton fields are transformed independently under the flavour group. When the gauge symmetry is broken by the Higgs mechanism, all of the particles acquire mass.  In the gauge basis the charged currents are diagonal, but  the  quark and lepton physical  masses are found by  diagonalization of their corresponding mass matrix.  This way, the charged currents $J^{\pm}_\mu$ are invariant under the transformations
of the flavour symmetry group if and only if  the up and down quark fields are transformed with the same flavour group.
Thus, the charged current invariance condition under the family symmetry group implies
that the up and down quark fields are transformed with the same group.

When the gauge symmetry is spontaneously broken, the quarks and leptons, as well as the gauge bosons $W^{\pm}$, $Z$ and the Higgs particle acquire
mass. Under exact $S(3)$ symmetry and only one $SU(2)$ Higgs doublet field, the mass spectrum for either up or down quark sectors consists of one massive particle (top and bottom quarks) in a singlet irreducible representation (irrep) and a pair of massless particles in a doublet irrep. The  $S(3)$ group treats three
objects symmetrically, while the hierarchical nature of the mass matrices is a consequence
of the representation structure $1 + 2$ of $S(3)$, which treats the generations differently. The $S(3)$ group is non-Abelian and has a doublet  and a singlet in its irrep. In order to properly generate non trivial quark and lepton mass matrices, the Higgs sector has to be flavour extended too, introducing three $SU(2)$ Higgs doublet fields $H_i, i=1,2,3$, which also transform as the irrep under $S(3)$~\cite{Kubo:2003iw}. That is, one Higgs field in a singlet and the other two in a
doublet irrep of $S(3)$.  The $S(3)$ flavour symmetry imposes some constrains on  the Higgs potential and in the end, after spontaneous symmetry breaking of the S(3)HDM, the $12 \times 12$ Higgs mass matrix can be exactly and analytically diagonalized, and
the physical Higgs masses become known as functions of  ten free parameters and the vacuum expectation values (vev's). The Higgs spectrum of the model consists of five neutral and four
charged Higgs bosons; three neutral Higgs bosons ($h_0$, $H_{1,2}$) are CP-even, whereas two neutral Higgs bosons ($A_{1,2}$) are CP-odd.

Up to now, the particle observed at  LHC  is a particle in the physical
spectrum of the Higgs boson of the SM. It is not known if there are one or many Higgs
bosons. An indication of the presence of one Higgs boson or an
extended Higgs sector, as the one proposed in the $S(3)$-invariant
extension of the Standard Model, could be found at the Large Hadron
Collider~\cite{Barger:2009me,Bhattacharyya:2010hp,Bhattacharyya:2012ze}.

A precise measurement of the trilinear Higgs self-couplings will also
make it possible to test extended Higgs models, which
have a different Higgs potential structure , and
hence different trilinear Higgs couplings as compared to SM. The $S(3)$ flavour group is the simplest flavour symmetry group; it imposes constrains that makes it even simpler than the 2HDM. In particular,
the model can easily accommodate all of the quark and lepton mixing matrices~\cite{RodriguezJauregui:1997pc,Mondragon:1998gy,Mondragon:1998yw,Mondragon:1999jt,Mondragon:2000ia,Canales:2013ura,Canales:2013cga,Hernandez:2014vta}.  In spite of the complicated nature of the Higgs potential, it is possible
to study in detail  the trilinear
couplings to the  CP-even, CP-odd and charged Higgs bosons in the S(3)HDM . 
 Thus, in the CP-conserving case, there are  allowed
trilinear Higgs scalar couplings which can be labelled as $\lambda_{ijk}$ with $i,\, j,\, k=h_0,\, H_1,\, H_2$,  all even in the number of $A$, involving the CP-odd Higgs boson. Thus, in the case of the S(3)HDM there are a total of  ten trilinear Higgs scalar couplings.

On the other hand, the introduction of additional symmetries to SM open up to understand the replication of families in that. Interesting extended models have been proposed in the literature (see for instance~\cite{Ma:2007ia} and
references therein, for a review on the subject). One flavour discrete symmetry such as $S(3)$ is  the minimal discrete group that reproduces the fermionic masses and mixing matrices~\cite{Derman:1978rx,Canales:2013ura,Canales:2013cga}. 

We study quantitatively the trilinear Higgs couplings, and compare these couplings to the corresponding Standard Model trilinear Higgs coupling in some regions of the parameter space. In particular, we are interested on the trilinear
($\lambda^{SM}_{HHH}$) and quartic ($\lambda^{SM}_{HHHH}$) self-couplings.
As we know, SM has just one trilinear self-coupling $\lambda^{SM}_{HHH}$, which is given by $\lambda^{SM}_{HHH} =3m^2_H/v$, where $v$ is the vacuum expectation value ($v=246\, \textrm{GeV}$). Writing the  free parameters as a function of the physical Higgs masses, the trilinear Higgs couplings of the model may be determined in terms of seven  Higgs mass eigenvalues, one  free parameter $\theta_S$  and the vev ratio $\tan \omega_3 \equiv {2v_2}/{v_3}$.

In this paper we investigate the trilinear
self-couplings of the $S(3)$ extended Standard Model, including an extended $S(3)$ Higgs boson sector. In the literature there are other extensions
of the SM, such as the  two Higgs doublet model (THDM)~\cite{Branco:2011iw} which also has 10 free parameters plus two vev's. 
In this  models  there are several trilinear Higgs couplings, with a more complicated dependence on the underlying masses. We assume that spontaneous $CP$ violation does not occur.

This paper is organized as follows: in section~\ref{sec:model} we present
some remarks about the $S(3)$ flavour symmetry and its Lagrangian, the Higgs sector of the $S(3)$ extended model, and the form of the Higgs mass matrix; in section~\ref{sec:3} we focus on the trilinear self-couplings of neutral Higgs bosons whereas the
details of the numerical results are presented 
in section~\ref{sec:4}, and finally  we present our
conclusions in section~\ref{sec:5}.

%%%%%%%%%%%%%%%%%%%%%%%%%%%%%%%%%%%%%%%%%%%%%%%%%%%%%%%%%%%%%%%%%%%%
\section{The $S(3)$ extended Higgs doublet model}\label{sec:model}
%%%%%%%%%%%%%%%%%%%%%%%%%%%%%%%%%%%%%%%%%%%%%%%%%%%%%%%%%%%%%%%%%%%%
The Lagrangian  ${\cal L}_{H}$ of the $S(3)$ extended Higgs sector incorporates three complex scalar SU(2) doublets fields. Such a theory is  based on purely aesthetic grounds: in view of the family replication of the elementary fermion spectrum one can speculate that this flavour symmetry is the symmetry of the fundamental particles,  and an  analogous flavour symmetry principle might work for the Higgs sector as well: 
\be
{\cal L}_{H_i} =\left[   D_\mu H_S\right]^2+\left[D_\mu H_1\right]^2
+\left[D_\mu H_2\right]^2-V\left( H_1,H_2,H_S \right),
\ee
where $D_\mu$ is the usual covariant derivative, the scalar potential
$V\left( H_1,H_2,H_S \right)$ is the most general Higgs potential
invariant under $SU(3)_C \times SU(2)_L \times U(1)_Y\times S(3)$~\cite{EmmanuelCosta:2007zz,Beltran:2009zz}, and in our case, the discrete flavour symmetry $S(3)$  Higgs doublets for this model are: 
\be
\ba{rcl} \label{eq:doubletshiggs}
H_1&=&\left( \ba{c}
\phi_1+i\phi_4\cr
\phi_7+i\phi_{10}
\ea
\right),\,
H_2=\left( \ba{c}
\phi_2+i\phi_5\cr
\phi_8+i\phi_{11}\ea
\right), \,
H_S=\left( \ba{c}
\phi_3+i\phi_6\cr
\phi_9+i\phi_{12}\ea
\right) .
\ea
\ee
The numbering of the real scalar $\phi_i$ fields is chosen for convenience when writing the mass matrices 
for the scalar particles, and the subscript $S$ is the flavour index for the Higgs field singlet under $S(3)$. $H_i$ with $i=1,2$ are the components of the $S(3)$ doublet field.
 In the analysis, it is better to introduce nine real quadratic forms $x_i$ invariant under $SU(2)\times U(1)$
\be\label{eq:xveccomp}
\ba{ccc}
x_1=H^\dagger_1 H_1,&
x_4= {\cal R}\left(H^\dagger_1 H_2\right),&
x_7= {\cal I}\left(H^\dagger_1 H_2\right),\cr
x_2=H^\dagger_2 H_2,&
x_5= {\cal R}\left(H^\dagger_1 H_S\right),&
x_8= {\cal I}\left(H^\dagger_1 H_S\right),\cr
x_3=H^\dagger_S H_S ,&
x_6= {\cal R}\left(H^\dagger_2 H_S\right),&
x_9={\cal I}\left(H^\dagger_2 H_S\right) .
\ea
\ee
Now, it is a simple matter to write down the potential
\be \label{eq:potential1}
\ba{lll}
V&=&\mu^2_1\left(x_1+x_2 \right)+\mu^2_0 x_3+ax^2_3+b\left(x_1+x_2 \right)x_3+c\left(x_1+x_2 \right)^2\\
&&-4dx_7^2+2e \left[ \left(x_1 - x_2\right)x_6+2x_4 x_5\right]+f\left(x_5^2+x_6^2+x_8^2+x_9^2 \right) \\
&&+g\left[\left(x_1-x_2 \right)^2+4x_4^2 \right]+2h\left(x_5^2+x_6^2-x_8^2-x_9^2 \right) .\\
\ea
\ee
where the $\mu_{0,1}^2$ parameters have dimensions of mass squared and the eight real couplings $a, \cdots, h$ are dimensionless free parameters.
The $S(3)$ invariant Higgs potential in Eq.~(\ref{eq:potential1})  has a minimum at  
\be\label{eq:minimo}
\phi_7={\it v_1},\ \phi_8={\it v_2},\ \phi_9={\it v_3},\
\phi_i=0, \,\,\,\, i\neq 7,8,9\, ,
\ee 
 where we have adopted for convenience vev's $v_i$ ($i=1,2,3$) which do not have any complex relative phase, $v_i \in \Re$. Such a minimum determines vector boson masses through the Higgs mechanism; in particular, for the charged gauge bosons  $W^{\pm}$ one obtains $m^2_{W}=(g^2/2) \sum_{i=1}^3 v^2_i$, with $g$ standing for the $SU(2)$ coupling constant. In a standard notation we can write $v_i=v\cos{w_i}$  where $v=246$ GeV is the electroweak scale and their $w_i's$ are three free parameters. Now we can  rewrite the potential $V$ and express it in a simple matrix form as
\be \label{eq:potential2}
V({\bf X} )={\bf A}^T{\bf X}+\frac{1}{2}{\bf X}^T{\bf B}{\bf X},
\ee
with the vector  ${\bf X}$ given by 

\be
{\bf X}^T=\left( x_1, x_2 ,x_3, \dots,x_9 \right),\label{appppa} 
\ee
${\bf A}$ is a mass parameter vector,
\begin{eqnarray}\label{eq:amatrix}
 {\bf A}^T=\left(\mu^2_1,\mu^2_1,\mu^2_0,0,0,0,0,0,0 \right)  \label{appppb}
\end{eqnarray}
and ${\bf B}$  is a $9\times9$ real parameter symmetric matrix
%
%\begin{widetext}
\be \label{eq:bmatrix}
{\bf B}= \left( \ba{ccccccccc}
2(c+g) & 2(c-g) &b&0&0&2e&0&0&0\cr
2(c-g) & 2(c+g) & b&0&0&-2e&0&0&0\cr
b&b&2a&0&0&0&0&0&0\cr
0&0&0&8g&4e&0&0&0&0\cr
0&0&0&4e&2(f+2h)&0&0&0&0\cr
2e&-2e&0&0&0&2(f+2h)&0&0&0\cr
0&0&0&0&0&0&-8d&0&0\cr
0&0&0&0&0&0&0&2(f-2h)&0\cr
0&0&0&0&0&0&0&0&2(f-2h)
\ea
\right).
\ee
%\end{widetext}
%
 The minimization conditions give us three equations  determined by demanding the vanishing of $\partial V/\partial \phi_i$.  Then we get the mass parameters $\mu_1$  and $\mu_0$  given as
\begin{equation}\label{eq:minimomuno}
\mu_1^2 =
-\displaystyle\left(b+f+2h\right)v_3^2-2\left(c+g\right)\left(v_1^2+v_2^2\right) +\frac{3e(v_1^2-2v_1v_2-v_2^2)v_3}{v_1-v_2} ,
\end{equation}
and
\begin{equation}\label{eq:minimomcero}
\mu_0^2 =
-\left[2av_3^2+\left(b+f+2h\right)\left(v_1^2+v_2^2\right) -e\left(\displaystyle\frac{3v_1^2-v_2^2}{v_3}\right)v_2
\right].
\end{equation}
 From these, the following relationship among the Higgs vev's is obtained:  
\begin{equation}\label{eq:minimovevs}
v_1=\sqrt{3} v_2 .
\end{equation}
This relationship between $v_1$ and  $v_2$ is similar to the condition imposed in Ref.~\cite{Canales:2013cga} to generate the Nearest Neighbour Interaction (NNI) mass matrices in the context of the SM $\otimes S(3)$ with two texture zeroes. Eqs.~(\ref{eq:minimomuno}-\ref{eq:minimovevs}) reduce the number of free  parameters  from thirteen to ten. In this model,
the Higgs boson masses  are obtained by diagonalizing the $12\times12$ mass matrix,
\be
(\mathcal{M}^2_H)_{ij} = \displaystyle\left.\frac{1}{2}\frac{\partial^2 V}{\partial\phi_i\partial\phi_j}\right|_{\hbox{mim}} ,
\ee
with $i,j=\overline{1,12}$. We have
\be\label{eq:matrizmasa}
\mathcal{ M}^2_H = \hbox{diag}\left({\bf M}_C^2, {\bf M}_C^2 , {\bf M}_S^2 , {\bf M}_P^2  \right) \, .
\ee
The $3\times 3$ symmetric and Hermitian sub-matrices $\bf {M}^2_C$ , $\bf {M}^2_S$, $\bf {M}^2_P$, are respectively the charged, scalar and pseudoscalar Higgs mass matrix. After diagonalizing the mass matrices, the masses of the physical charged, scalar  and pseudoscalar Higgs bosons are obtained. From the minimization condition (\ref{eq:minimovevs}), expressing the vev's of the Higgs fields as $v_i=v\cos\omega_i$ and the relationship  $v^2=\sum_{i=1}^3 v^2_i$,  the mass matrix $\mathcal{M}^2_H $ can be parametrized with eight  free parameters and  $\omega_3$ given as
\be\label{eq:omega3}
\tan\omega_3= \frac{2v_2}{v_3}
\ee
where $\sin \omega_3 = 2v_2/v$ and $\cos \omega_3 = v_3/v$.

%%%%%%%%%%%%%%%%%%%%%
\subsection{S(3)-Higgs mass matrices}
%%%%%%%%%%%%%%%%%%%%%
The S(3)HDM involves twelve independent scalar fields; three of them can be identified with the would be Goldstone bosons $W^{\pm}$, $Z$ and the remaining nine correspond to physical Higgs particles. The proper Goldstone and Higgs fields are found through a diagonalization of the Higgs mass matrix (ref{}). The Higgs boson masses in this model are obtained by diagonalizing the exact and explicit $12\times12$ mass matrix in Eq.~(\ref{eq:matrizmasa}),
\bea\label{eq:charged}
{M}_C^2 = \left(
\ba{ccc}
c_{11} & c_{12} & c_{13} \\
 & c_{22} & c_{23} \\
 &  & c_{33} 
\ea \right) ,\qquad {M}_S^2 = \left(
\ba{ccc}
s_{11} & s_{12} & s_{13} \\
 & s_{22} & s_{23} \\
 &  & s_{33} 
\ea \right) ,\qquad {M}_P^2 = \left(
\ba{ccc}
p_{11} & p_{12} & p_{13} \\
 & p_{22} & p_{23} \\
 &  & p_{33} 
\ea \right) 
\eea
with the following matrix entries
\be
\ba{l}
c_{11} = -4gv_2^2-(4ev_2+f'v_3)v_3 \\
c_{12} =  2\sqrt{3}(2gv_2+ev_3)v_2 \\
c_{13} =  \sqrt{3}\left(2ev_2+f'v_3\right)v_2 \\
c_{22} = -12gv_2^2-\left(8ev_2+f'v_3\right)v_3 \\
c_{23} = \left(2ev_2 +f'v_3\right)v_2 \\
c_{33} = -\displaystyle{\frac{4v_3}{v_2^2}}\left(2ev_2+f'v_3\right) .
\ea
\ba{l}
s_{11} =  12(c+g)v_2^2  \\
s_{12} =  2\sqrt{3}  \left( 2(c+g)v_2 + 3ev_3\right)v_2  \\
s_{13} =  2\sqrt{3}  \left( 3ev_2 + (b+f') v_3\right)v_2 \\
s_{22} =  4\left((c+g)v_2 -3ev_3 \right)v_2 \\
s_{23} =  2 \left( 3 e v_2 + (b+f')v_3\right)v_2 \\
s_{33} =  -\displaystyle\frac{8ev_2^2}{v_3}+4av_3^2 .
\ea
\ba{l}
p_{11} =  -4\left((d+g)v_2^2+ev_2v_3+hv_3^2\right)   \\
p_{12} =  2\sqrt{3}  \left(2(d+g)v_2+ ev_3\right)v_2   \\
p_{13} =  2\sqrt{3} (ev_2+2hv_3)v_2 \\
p_{22} =  -12(d+g)v_2^2-8ev_2v_3 -4hv_3^2 \\
p_{23} =  2(ev_2+2hv_3)v_2  \\
p_{33} =  -\displaystyle{\frac{8 v_2^2}{v_3}}(ev_2+2hv_3) .
\ea
\label{eq:entriesmasses}
\ee
where $f'\equiv f+2h$.The physical Higgs masses are found from the diagonalization process
\be\label{eq:mdiag}
[{\mathcal{M}}_{diag}^2 ]_{i}= R^T_{i}  {\mathcal{M}}_{i}^2 R_{i}\qquad i={c,s,p}
\ee
where the indices $c,s,p$ stand for charged, scalar and pseudoscalar respectively. Then,  the rotation matrices $R_i$  are
\be\label{eq:roteven}
R_{i} =  \left(
\begin{array}{ccc}
-1& 0 & 0 \\
0 & 1 & 0 \\
0 & 0 &-1
\end{array}\right)
\left(
\begin{array}{ccc}
1/2 & -\sqrt{3}/2 & 0 \\
\sqrt{3}/2 & 1/2 & 0  \\
0 & 0 &1
\end{array}\right)
\cdot Q_{i} ,\qquad i={c,s,p}
\ee
the mixing matrix  $ Q_{i}$ is 
\be
Q_{i} =  \left(
\begin{array}{ccc}
1 & 0 & 0 \\
0 & {\cos{\theta}}_{i} &{\sin{\theta}}_{i}  \\
0 & -{\sin{\theta}}_{i} &{\cos{\theta}}_{i}
\end{array}\right)\qquad i={C,S,P}
\label{eq:rmatrix}
\ee
with
\be\label{eq:thetai}
\qquad\tan{\theta_S}=\displaystyle \frac{2{s_{13}}}{\sqrt{3}(m_{H_1}^2-{s_{33}})},
\qquad\tan{\theta_C}=\tan{\theta_P}=\frac{1}{\tan{\omega_3}}.
\ee

We start by considering the Higgs  mass matrix ${\bf M}_s^2$  for the CP-even Higgs. Defining the physical mass eigenstates $m_{h_0}^2$, $m_{H_1}^2$, and $m^2_{H_2}$, the physical Higgs masses are found from the diagonalization process.The physical masses for the CP-even Higgs boson scalars are:
\be
\ba{rcl}
m_{h_0}^2 &=& -18ev_2v_3 \cr
m_{H_1,H_2} ^2 &=& (\M_a^2 + \M_c^2) \pm \sqrt{(\M_a^2 - \M_c^2)^2 + (\M_b^2)^2},
\ea
\ee
where
\be
\ba{l}
\M_a^2=v_2(8(c+g)v_2+3ev_3) \\
\M_b^2=4v_2\left(3ev_2+(b+f+2h)v_3\right) \\
\M_c^2=-\displaystyle\frac{4ev_2^3}{v_3}+2av_3^2 ,
\ea
\ee
The scalar Higgs mass matrix  ${\cal{M}}^2_{s} $ is diagonalized by a  rotation matrix $\bf R_s$ given by  Eq.~({\ref{eq:rmatrix}}) which is   parametrized with
\be\label{eq:theta}
\tan{\theta_S}=\displaystyle \frac{2{s_{13}}}{\sqrt{3}(m_{H_1}^2-{s_{33}})} =\displaystyle\frac{\M_b^2}{2\M_a^2-m_{H_2}^2} 
\ee
where ${s_{13}}$ and ${s_{33}}$ are obtained from Eq.~(\ref{eq:entriesmasses}). 
The charged Higgs mass matrix  ${\cal{M}}^2_{c} $ is diagonalized by a  rotation matrix $\bf R_c$ in  Eq.~({\ref{eq:rmatrix}}), and the charged Higgs boson masses are
$$
m_{H_1^\pm}^2 = -\left(10ev_2+(f+2h)v_3\right)v_3 -16gv^2_2,
\qquad
m_{H_2^\pm}^2 = -\displaystyle\frac{v^2}{v_3}\left(2ev_2+(f+2h)v_3\right) .
$$

The resulting squared masses for the CP-odd  pseudoscalar bosons Higgs are 
\be
\ba{rcl}
m_{A_1}^2&=&-16 (d+g) v_2^2-10 e v_2v_3-4 h v_3^2, \cr
m_{A_2}^2&=&-\displaystyle\frac{2 (e v_2+2 h v_3)(4 v_2^2+v_3^2) }{v_3}.
\ea
\ee

Of the original twelve scalar degrees of freedom, three Goldstone bosons ($G^\pm$ and $G^0$) are absorbed by $W^\pm$ and $Z$. The remaining nine physical Higgs particles are three $CP-$even scalars ($h_0$ and $H_1$, $H_2$, with $m_{h_0}\leq m_{H_1}\leq m_{H_2}$), two $CP$-odd scalars ($A_1$, and $A_2$, with $m_{A_1}\leq m_{A_2}$), and two charged Higgs pairs ($H^\pm_{1,2}$, mass degenerate).  

%%%%%%%%%%%%%%%%%%%%%%%%%%
\subsection{The S(3)HDM free parameter space}
%%%%%%%%%%%%%%%%%%%%%%%%%%

The  Higgs mass matrix $ M^2_H$ contains ten real free independent parameters, 
namely, $ \mu_0$, $ \mu_1$ , eight  dimensionless parameters $a,b,c,d,e,f,g,h$, and the three vacuum expectation values $v_1=v\cos\omega_1 $, $v_2=v\cos\omega_2 $,
 $v_3=v\cos\omega_3$ . These thirteen free parameters 
may be readily expressed in terms two free parameters and  seven Higgs mass eigenvalues as follows: first, we use the minimization conditions to eliminate  $ \mu_0$, $ \mu_1$ and $v_1$,  it reduces the number of free parameters from thirteen to ten, we then use the vev  relation $v^2_1+v^2_2+v^2_3=v^2=246\, \textrm{GeV}$ and the seven squared Higgs mass eigenvalue equations. We end up with two free parameters  namely  the vev $\omega_3$  and the scalar Higgs mixing angle $\theta_S$.

\begin{figure}[h] 
\centering 
\includegraphics[scale=1.0]{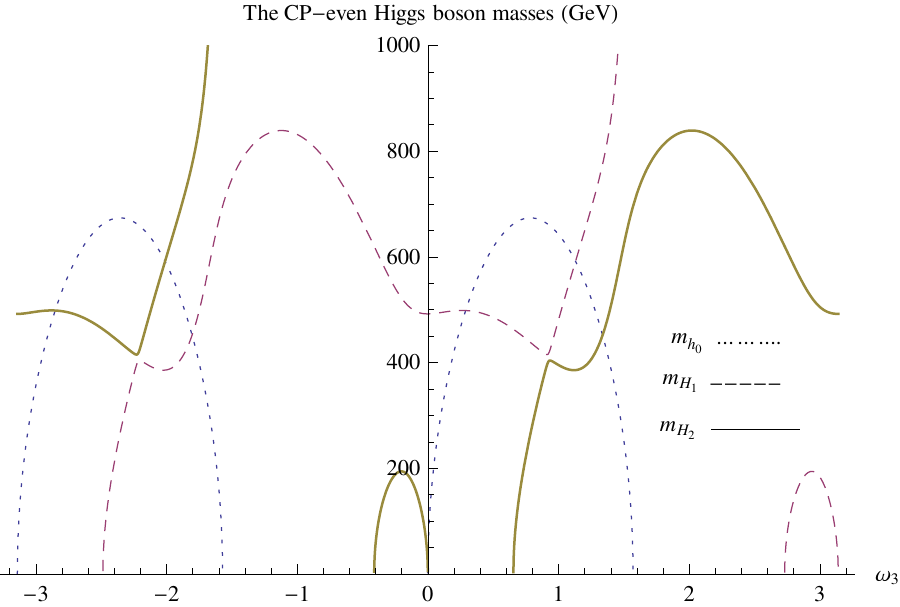} 
\caption{The masses of three CP-even Higgs scalars with respect to $\omega_3$, $-\pi\leq\omega_3\leq\pi$. The dashed line is for  $H_1$,  the solid line is for $H_2$ and the dotted line is for $h_0$.}
\label{fig:uno} 
\end{figure}

Higgs boson masses are not determined {\it a priori} within the theory and their decay patterns depend strongly on the masses. In order to determine the possible decay modes and branching ratios, it is necessary to investigate the change of mass spectrum with respect to the quadrilinear $a, \cdots, h$ couplings. For simplicity, we assume in our numerical analysis  the following values for these dimensionless parameters
$
a = 1,\ \  b = 1,\ \  c = 1,\ \  d = -1,\ \  e = -1,\ \  f = 3/2,\ \  g = 1,\ \  h = 1/2,
$
which guarantees that matrix  $B$, Eq.~(\ref{eq:bmatrix}), will be positive definite, and $v_2 = 246 (\sin\omega_3)/2$ GeV, $v_3 = 246 \cos\omega_3$ GeV,  $-\pi \le \omega_3 \le \pi$. Figure \ref{fig:uno} shows the masses of three CP-even Higgs scalars with respect to $\omega_3$.
The quadrilinear $a, \dots, h$ couplings are functions of $\theta_S$ and $w_3$, and are related to the masses of the Higgs bosons by
\be
\ba{l}
a=\displaystyle\frac{9\M_c^2v_3^2-2m_{h_0}^2v_2^2}{18v_3^4} \\
b=\displaystyle\frac{2m_{h_0}^2v^2+9\M_b^2v_3}{36v_2v_3^2}+\frac{m_{H_2^\pm}^2}{v^2} \\
c=\displaystyle\frac{m_{H_2^\pm}^2}{4v^2}-\frac{m_{h_0}^2-9(m_{H_1^\pm}^2-m_{H_2^\pm}^2+2\M_a^2)}{144v_2^2} \\
d=\displaystyle\frac{1}{16}\left(\frac{m_{A_2}^2-m_{A_1}^2+m_{H_1^\pm}^2-m_{H_2^\pm}^2}{v_2^2}-\frac{m_{A_2}^2-m_{H_2^\pm}^2}{4v^2} \right)\\
e=\displaystyle-\frac{m_{h_0}^2}{(18v_2v_3)} \\
f=\displaystyle\frac{1}{18}\left(\frac{9(m_{A_2}^2-2m_{H_2^\pm}^2)}{v^2}+\frac{m_{h_0}^2}{v_3^2}\right) \\
g=\displaystyle\frac{4m_{h_0}^2-9(m_{H_1^\pm}^2-m_{H_2^\pm}^2)}{144v_2^2}-\frac{m_{H_2^\pm}^2}{4v^2} \\
h=\displaystyle\frac{1}{36}\left(\displaystyle\frac{m_{h_0}^2}{v_3^2}-\frac{9m_{A_2}^2}{v^2}\right) .
\ea
\label{eq:parameters}
\ee
where
\be
\ba{l}
\M_a^2 = \displaystyle\frac{1}{4}\left( m_{H_1}^2 +
       m_{H_2}^2 - (m_{H_1}^2 - m_{H_2}^2)\cos{2 \theta_S}    \right) \\
\\
\M_b^2= (\M_a^2-\M_c^2)\tan2\theta_S \\
\\
\M_c^2= \displaystyle\frac{1}{4}  
  \left( m_{H_1}^2 +
       m_{H_2}^2  +(m_{H_1}^2 - m_{H_2}^2)\cos2 \theta_S \right) .
\ea
\ee
\begin{figure}[h] 
\centering 
\includegraphics[scale=1]{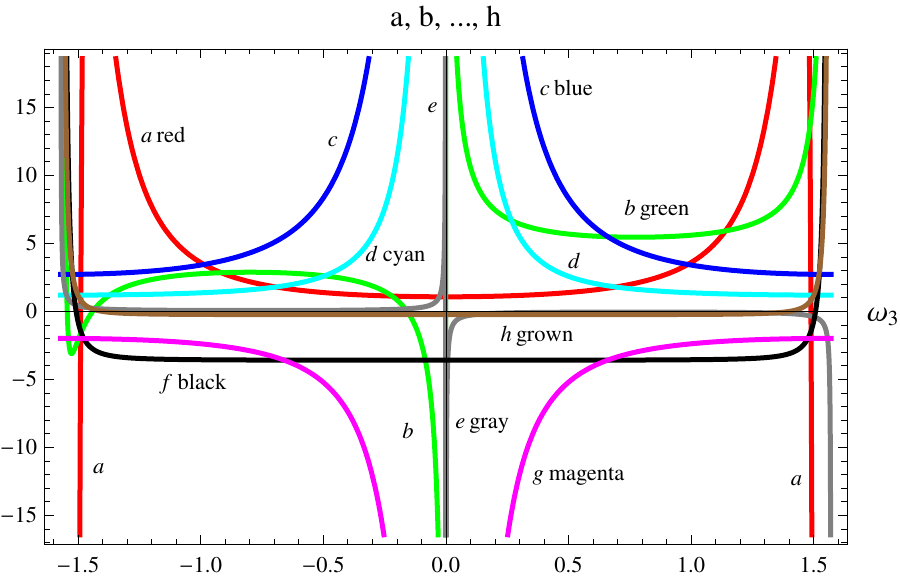} 
\caption{The quadrilinear $a, \dots, h$ couplings with respect to $\omega_3$, in the range $-\pi/2\leq\omega_3\leq\pi/2$.  The $\omega_3$ values close to zero are on the order of tenths, $\tan\omega_3\sim 0.09$. When $\omega_3$ approaches to $\pi/2$,  $\tan\omega_3$ diverges. Higher values are acceptable for $\tan\omega_3=5$. If $\tan\omega_3=1$, and then the mass spectrum of bosons Higgs is: $m_{h_0}= 677$ GeV, $m_{H_1}= 516$ GeV, $m_{H_2}= 125$ GeV, $m_{A_1}= 879$ GeV, $m_{A_2}= 248$ GeV, $m_{H_1^\pm}= 175$ GeV, $m_{H_2^\pm}= 350$ GeV. 
Where $a \sim 1$, $b \sim 0.5$, $c \sim 1.3$, $d \sim -5.4$, $e \sim -1.7$, $f \sim 3.4$, $g \sim 1$, $h \sim 0.2$.}
\label{fig:quadrilinear} 
\end{figure} 
Figure \ref{fig:quadrilinear}  shows the behavior of the quadrilinear couplings in Eq.~(\ref{eq:parameters}) where we have taken $\theta_S = \pi/3$ and $\omega_3$ as a free parameter, Eq.~(\ref{eq:omega3}). It is possible to specify a value for $\omega_3$ and from this, determine a mass spectrum of Higgs bosons. For example, for $\omega_3$ = 1, quadrilinear couplings take the values: $
a = 5.33,\ \  b = 0.16,\ \  c = 0.78,
d = -0.22,\ \  e = -1.94,\ \  f = 3.28,\ \  g = 0.93,\ \  h = 0.46,
$
fixing the Higgs mass spectrum to
$
m_{h_0}=693 \  \hbox{\rm  GeV} , \quad m_{H_1}=790 \   \hbox{\rm  GeV}, \quad m_{H_2}=125 \   \hbox{\rm  GeV},
m_{H_1^\pm}=181 \   \hbox{\rm  GeV} , \quad m_{H_2^\pm}=268 \    \hbox{\rm  GeV} ,
m_{A_1}=334  \   \hbox{\rm  GeV} ,\quad m_{A_2}=265 \   \hbox{\rm  GeV} .
$

%%%%%%%%%%%%%%%%%%%%%%%%%%%%%%%%%%%
\section{Trilinear self-couplings of neutral Higgs bosons}\label{sec:3}
%%%%%%%%%%%%%%%%%%%%%%%%%%%%%%%%%%%
The measurement of the Higgs self-coupling is crucial to determine the Higgs potential. Self-couplings are uniquely determined in the SM by the mass of the Higgs boson, which is related to the quadrilinear coupling $\lambda$ through $M_H=\sqrt{2\lambda v}$. The trilinear and quadrilinear vertices of the Higgs field $H$ are given by the coefficients: 
\be
\lambda_{HHH}= \lambda v =\displaystyle \frac{M_H^2}{2v}, \qquad \lambda_{HHHH} = \frac{\lambda}{4} = \frac{M_H^2}{8v^2}.
\ee
 The following definitions are often used:
\be\label{eq:lambdaijk}
\lambda_{ijk}=\frac{-i\partial^3V}{\partial H_i\partial H_j \partial H_k},
\ee
which are most easily obtained from the corresponding derivatives of $V$ in 
Eq.~(\ref{eq:potential1}) with respect to the fields $\{\phi_i\}$ with $i=1,\cdots,12$.
We can then write the trilinear couplings in terms of the derivatives of the potential (\ref{eq:potential1}) with respect to $\phi_i$ and the elements of the rotation matrix $R$ Eq.~(\ref{eq:roteven}) as
\be
 \lambda_{ijk}=N\sum_{lmn}R_{il}R_{jm}R_{kn} \frac{\partial^3V}{\partial \phi_i\partial \phi_j \partial \phi_k},
\ee
where the indices $l$, $m$, $n$ refer to the weak field basis, and $l\leq m\leq n=1,2,3$, $N$ is a factor of $n!$ for $n$ identical fields. We now proceed to obtain these couplings in an explicit form. The trilinear self-couplings $a_{lmn}=\displaystyle\frac{\partial^3V}{\partial \phi_i\partial \phi_j \partial \phi_k}$ among the neutral Higgs bosons can be written as 
\be
\begin{array}{ll}
a_{1,1,1}=6 \sqrt{6} (c+g) v_2 , & a_{1,1,2} , =\sqrt{2} (2 (c+g) v_2+3 e v_3) , \\
a_{1,1,3}=\sqrt{2} (3 e v_2+(b+f+2 h) v_3) , & a_{1,2,2}=2 \sqrt{6} (c+g) v_2 , \\
a_{1,2,3}=3 \sqrt{6} e v_2 , & a_{1,3,3}=\sqrt{6} (b+f+2 h) v_2 , \\
a_{2,2,2}=3 \sqrt{2} (2 (c+g) v_2-e v_3) , & a_{2,2,3}=\sqrt{2} ((b+f+2 h) v_3-3 e v_2) ,\\
a_{2,3,3}=\sqrt{2} (b+f+2 h) v_2 , & a_{3,3,3}=6 \sqrt{2} a v_3 .\\
\end{array}
\ee
Then, here we have ten trilinear Higgs scalar self-couplings and substituting for the elements of the rotation matrix, Eq.~(\ref{eq:roteven}), one obtains 
\be 
\ba{ll}
\lambda_{1,1,1} = 6v \left(\lambda_1s\omega_3 + \lambda_2 c\omega_3 \right) , & 
\lambda_{2,2,2} = 6v \left(\lambda_3s\omega_3 + \lambda_4 c\omega_3 \right) ,\\ 
\lambda_{3,3,3} = 6v \left(\lambda_5s\omega_3 + \lambda_6 c\omega_3 \right) , & 
\lambda_{1,1,2} = 2v \left(\lambda_7s\omega_3 + \lambda_8 c\omega_3 \right) , \\ 
\lambda_{1,1,3} = 2v \left(\lambda_9s\omega_3 + \lambda_{10} c\omega_3 \right) ,&
\lambda_{1,2,2} = 2v \left(\lambda_{11}s\omega_3 + \lambda_{12} c\omega_3 \right) , \\ 
\lambda_{1,2,3} = v \left(\lambda_{13}s\omega_3 + \lambda_{14} c\omega_3 \right) , &
\lambda_{1,3,3} = 2v \left(\lambda_{15}s\omega_3 + \lambda_{16} c\omega_3 \right) , \\ 
\lambda_{2,2,3} = 2v \left(\lambda_{17}s\omega_3 + \lambda_{18} c\omega_3 \right) , & 
\lambda_{2,3,3} = 2v \left(\lambda_{19}s\omega_3 + \lambda_{20} c\omega_3 \right) \, . 
 \ea
 \ee
$\lambda_1, \dots, \lambda_{20}$ depend on the quadrilinear coupling parameters of the Higgs potential Eq.~(\ref{eq:potential1}) and the mixing angle $\theta_S$  Eq.~(\ref{eq:theta}). For example, one can use the results of the appendix to compute the first $\lambda_{h_0h_0h_0}=\lambda_{1,1,1}$, these results are given in  Appendix \ref{Ap:apendixea},
\be 
\ba{rl}
\lambda_{h_0h_0h_0} &= \frac{\sqrt{3}}{4}v\left[2 (18as_{\theta_S}^3+  (b + f + 2 h)(3c_{\theta_S}^2+1)s_{\theta_S} -9e c_{\theta_S}^3 +3e c_{\theta_S})c\omega_3 \right. \\
 & \left. +( 3 (b + f + 2 h) (c_{\theta_S}-1) s_{\theta_S}^2 +  2 (c + g) ) 
 (9 c_{\theta_S}^3 - 3 c_{\theta_S}^2 + c_{\theta_S} - 3) \right. \\
 & \left.  - 3 e (3 c_{\theta_S} (c_{\theta_S}+ 1) - 1)  s_{\theta_S} ) s \omega_3 \right] .
\ea
\ee
Where $s_{\theta_S} = \sin\theta_S$, $c_{\theta_S} = \cos\theta_S$, $s\omega_3 = \sin\omega_3$, and $c{\omega_3} = \cos\omega_3$
\begin{figure}[h] 
\centering 
\includegraphics[scale=.6]{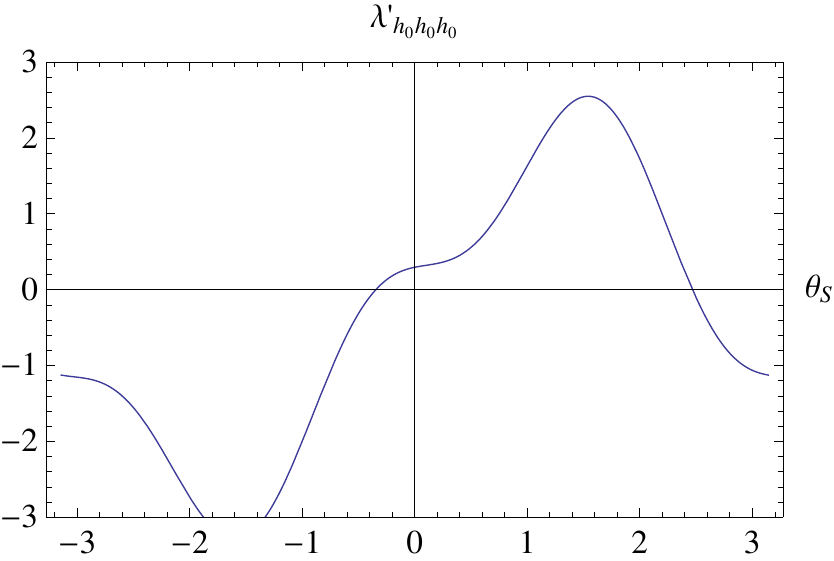}
\includegraphics[scale=.6]{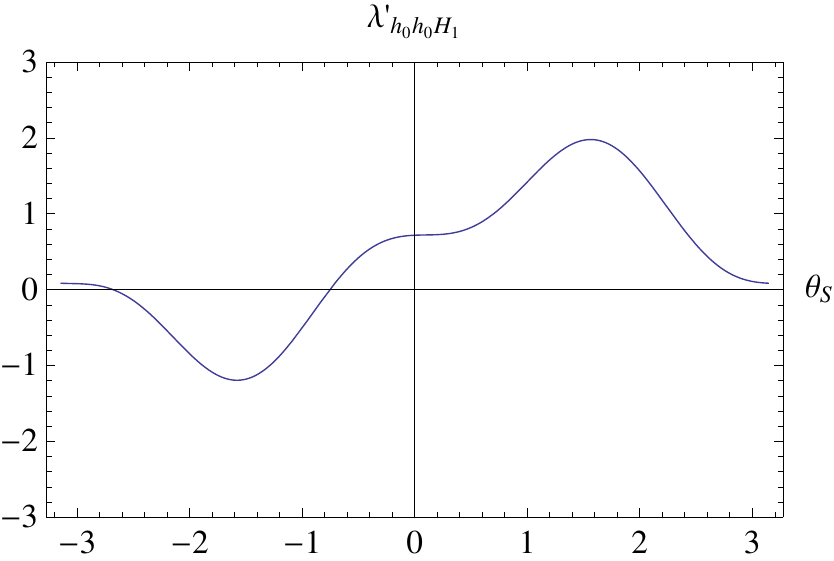}
\includegraphics[scale=.6]{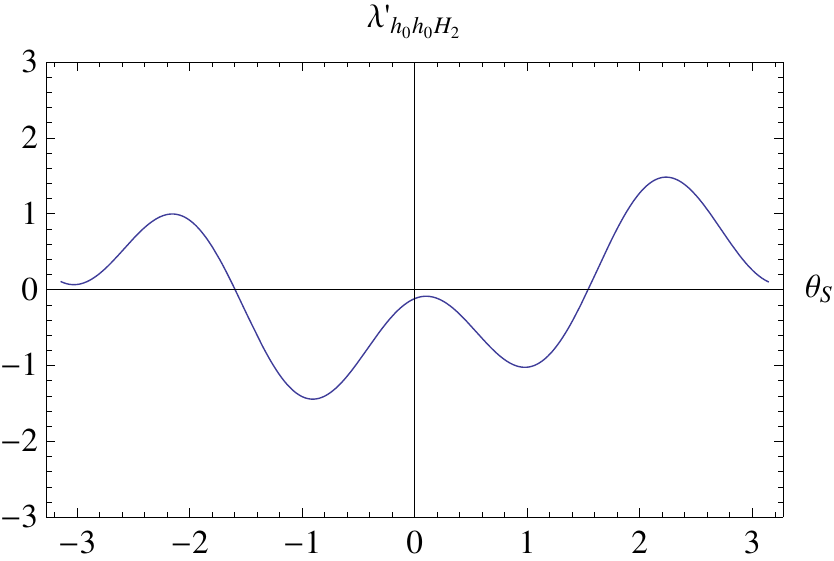}
\includegraphics[scale=.65]{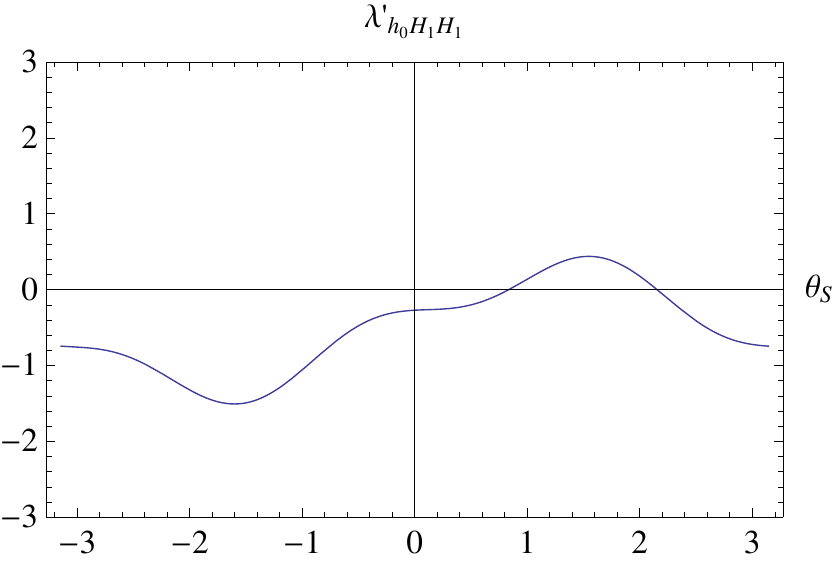}
\includegraphics[scale=.65]{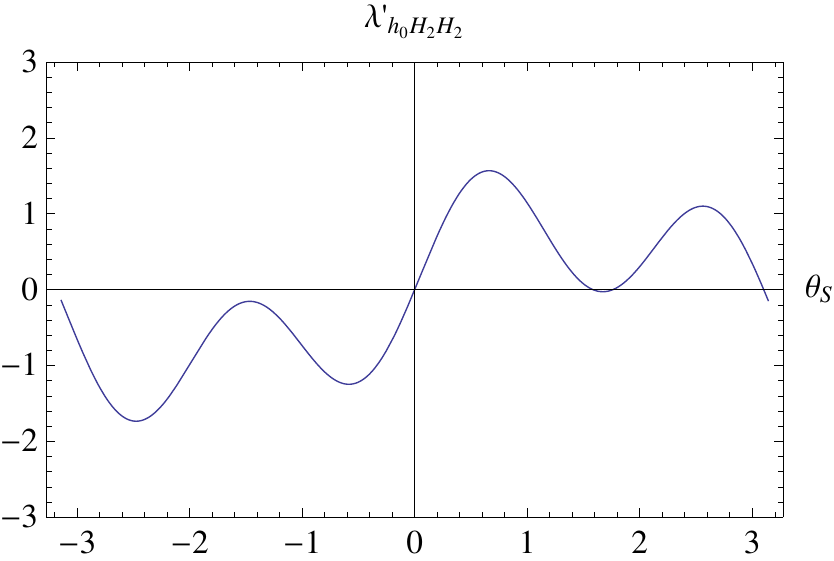}
\includegraphics[scale=.65]{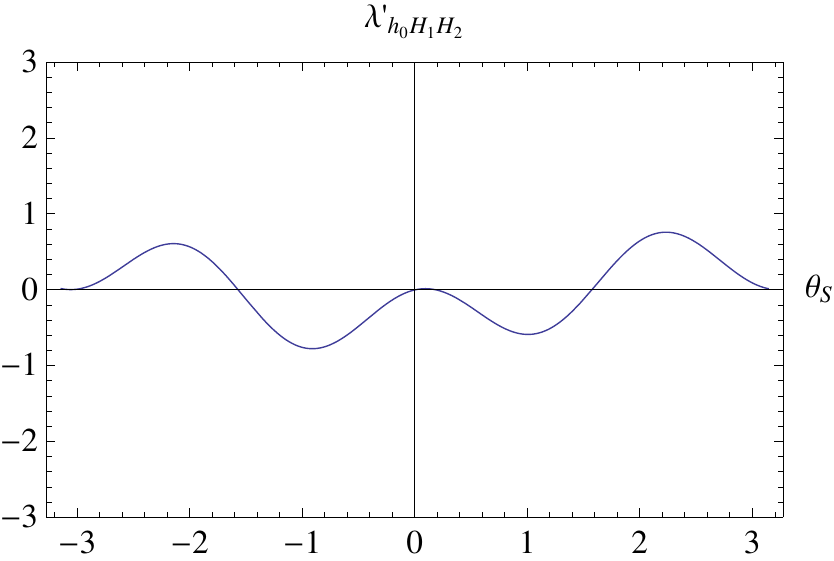}
\includegraphics[scale=.65]{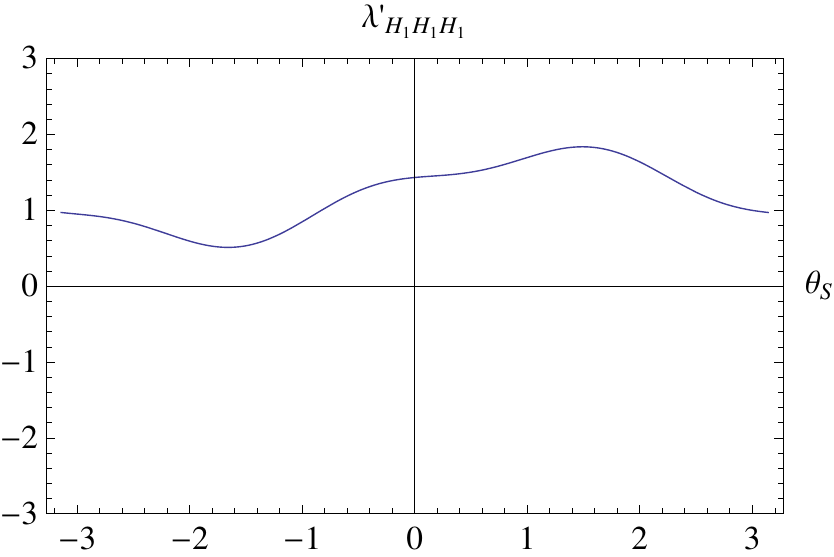}
\includegraphics[scale=.65]{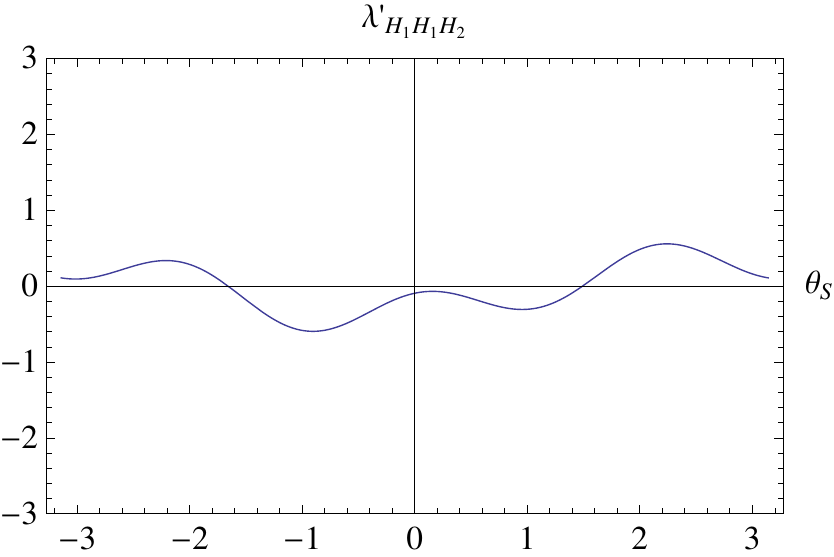}
\includegraphics[scale=.65]{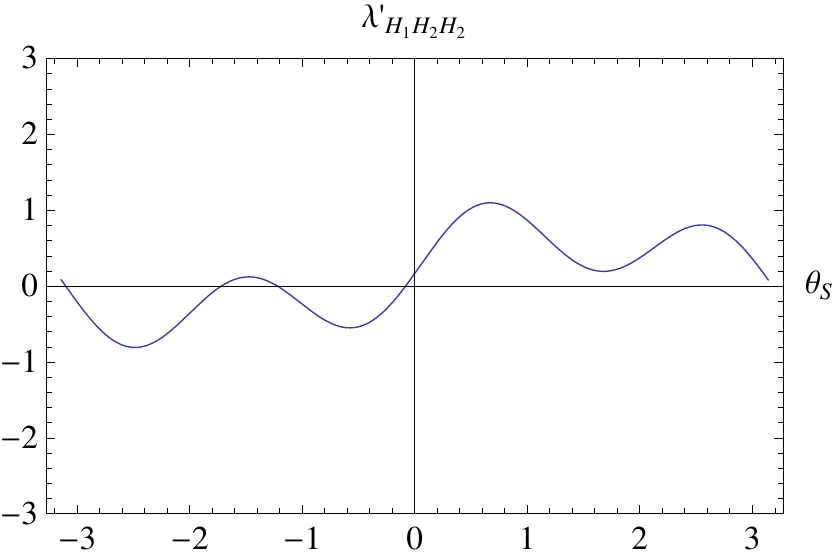}
\includegraphics[scale=.65]{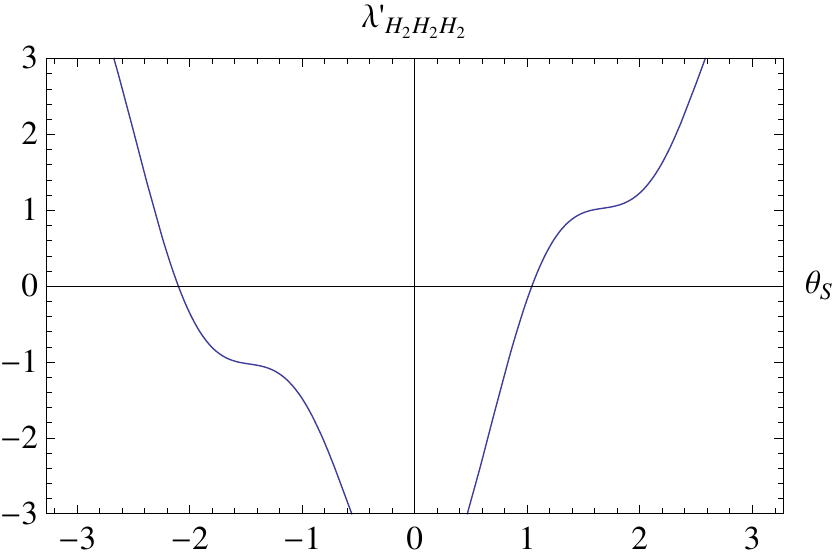}
\caption{The trilinear couplings by CP-even Higgs bosons, $\lambda_{ijk}'=\lambda_{ijk}/(125\ \hbox{\rm GeV})^2$ with $i$, $j$, $k$ = $h_0$, $H_{1,2}$, as a function of the mixing angle $-\pi\leq\theta_S\leq\pi$, Eq.~(\ref{eq:theta}), where $\tan\omega_3=1$.}  \label{fig:hcpevenmas}
\end{figure} 

%%%%%%%%%%%%%%%%%%%%
\section{Numerical results}\label{sec:4}
\begin{figure}[h] 
\centering 
\includegraphics[scale=.5]{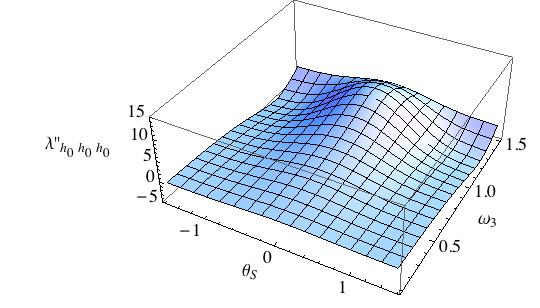} 
\includegraphics[scale=.5]{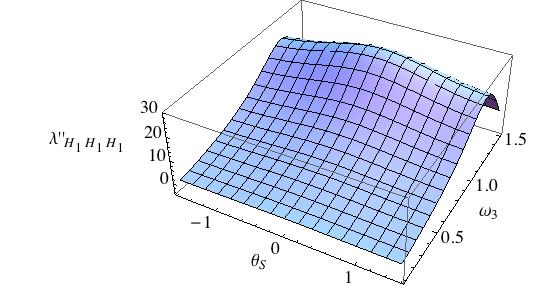}
\includegraphics[scale=.5]{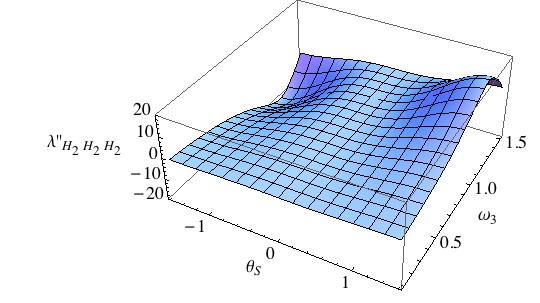}
\caption{The trilinear couplings by CP-even Higgs bosons, $\lambda_{ijk}''=\lambda_{ijk}/(m_{H_2})^2$ with $i$, $j$, $k$ = $h_0$, $H_{1,2}$, as a function of the mixing angle $-\pi/2\leq\theta_S\leq\pi/2$, and $0.1 \leq\omega_3\leq 1.5$.}  
\label{fig:3dtmas}
\end{figure}

%%%%%%%%%%%%%%%%%%%%
Here, we present numerical results considering that  one of the CP-even Higgs scalar boson behaves like the Higgs boson of the Standard Model, where the mass of the lightest Higgs scalar is significantly smaller than the masses of the other six Higgs bosons of the model. We assume that it is light and its mass is around 125 GeV. In our analysis we propose  $H_2$ as our candidate. The other Higgs bosons have consequently masses  of the order of  $\sim 500$ GeV and without loss of generality and for the sake of simplicity, we consider that their masses are not equal. Figure \ref{fig:hcpevenmas} shows the trilinear couplings by CP-even Higgs bosons as a function of the mixing angle $\theta_S$, Eq.~(\ref{eq:theta}),  where we have fixed $\tan\omega_3=1$. The higher strength coupling corresponds to $\lambda_{h_0h_0h_0}$ and $\lambda_{H_2H_2H_2}$, whereas that $\lambda_{H_1H_1H_1}$ became very small,
$$
\lambda_{h_0h_0h_0} \sim \lambda_{H_2H_2H_2} > \lambda_{H_1H_1H_1},
$$
 followed by $\lambda_{h_0h_0H_2}$, $\lambda_{h_0h_0H_1}$ ($\lambda_{h_0H_2H_2}$) and others. We proposed that the lightest is $H_1$ and its trilinear coupling has the lowest intensity. Although there are couplings equal in shape, they are different in intensity. Comparing these results only with the SM trilinear coupling, we see that our values are within acceptance ranges in the literature, but they are richer for a Higgs potential analysis, enabling better understanding of the spontaneous weak symmetry breaking.
\begin{figure}[h] 
\centering 
\includegraphics[scale=.6]{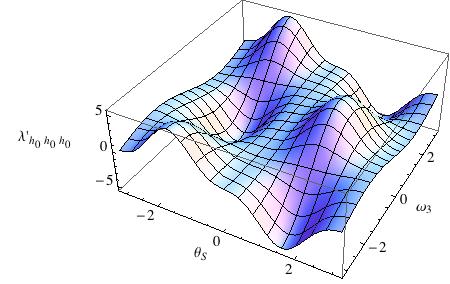}
\includegraphics[scale=.6]{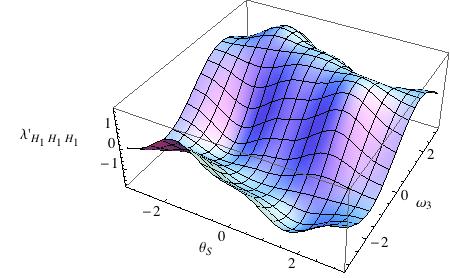}
\includegraphics[scale=.6]{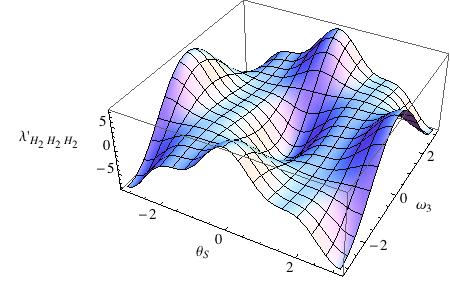}
\caption{The trilinear couplings by CP-even Higgs bosons, $\lambda_{ijk}'=\lambda_{ijk}/(125\ \hbox{\rm GeV})^2$ with $i,j$, $k=h, \, H_{1,2}$, as a function of the mixing angle $-\pi\leq\theta_S\leq\pi$, and $-\pi\leq\omega_3\leq\pi$.}  
\label{fig:3dtw}
\end{figure} 
Furthermore, when comparing the self-couplings of the Higgs boson as a function of the two mixing angles $\theta_S$ and $\omega_3$, there is a symmetry, Figure \ref{fig:3dtw}. This allows us to characterize different Higgs bosons. The mass of the Higgs boson is fixed through $\omega_3$ and couplings are also determined  by mixing Higgs bosons depending on $\theta_S$. Our results are normalized ($M_h=125$ GeV) in Figure \ref{fig:hcpevenmas}, but with either parametrization three scalar bosons dependence is observed. See figure \ref{fig:3dtw}. We can analyze the behavior of the couplings, depending on the mass of the boson $M_{H_2}$ through $\omega_3$, see Figure \ref{fig:3dtmas}.
%

%%%%%%%%%%%%%%%%%%
\section{Conclusions}\label{sec:5}
%%%%%%%%%%%%%%%%%%
We studied only the scalar sector assuming the pseudoscalars to be too heavy to be relevant. In this work we have analyzed the complete scalar sector of  an $S(3)$ flavour model. We deal with three $CP$-even, two $CP$-odd and two sets of charged scalar particles.  We have improved our potential minimization technique which enabled us to explore a larger region of the allowed 
parameter space. We have studied in detail the trilinear couplings of the lightest Higgs boson of this model.
Within the allowed domain of the parameter space of the model, the trilinear Higgs  couplings  have a  strong dependence on  $\tan\omega_3=2v_2/v_3$ and $\tan\theta_S$. The extended Higgs spectrum in $S(3)$ models gives rise to numbers of trilinear couplings. The $hhh$ coupling can be measured in $hh$ continuum production linear colliders as at $e^+e^-$.

\appendix
%%%%%%%%%%%%%%%%%%%
\section{Trilinear couplings $\lambda_{i,j,k}$} \label{Ap:apendixea}
%%%%%%%%%%%%%%%%%%%
For completeness, we show the explicit neutral scalar Higgs bosons self-couplings.
\be
\ba{rl}
\lambda_{h_0h_0h_0} &= 6v \left(\lambda_1s\omega_3 + \lambda_2 c\omega_3 \right) , \\
\lambda_1 &= \displaystyle\frac{3\sqrt{3}}{4} (b + f + 2 h) (c_{\theta_S} - 1) s_{\theta_S}^2  \\
\lambda_2 &=  \displaystyle\frac{2\sqrt{3}}{4} \left(18 a s_{\theta_S}^3 +\left[  (b + f + 2 h)s_{\theta_S} - 3ec_{\theta_S}\right] (3 c_{\theta_S}^2 + 1)\right)\\
 & +  2 (c + g)  (9 c_{\theta_S}^3 - 3 c_{\theta_S}^2 + c_{\theta_S} - 3) - 3 e (3 c_{\theta_S} (c_{\theta_S}+ 1) - 1)  s_{\theta_S} \\
%\ea
%\ee
%\be 
%\ba{rl}
\lambda_{H_1H_1H_1} &= 6v \left(\lambda_3s\omega_3 + \lambda_4 c\omega_3 \right) ,\\ 
 \lambda_3 &= \displaystyle\frac{1}{4}\left[ (b + f + 2 h) (c_{\theta_S}  + 3) s_{\theta_S}^2 + 
  6 (c + g) (c_{\theta_S}^3 + c_{\theta_S}^2 + c_{\theta_S}  + 9) \right. \\
& \left.  - 3 e ((c_{\theta_S}  - 3) c_{\theta_S}  - 3) s_{\theta_S} \right] \\
\lambda_4 &=\displaystyle\frac{1}{2} \left[6 a s_{\theta_S}^3 + (b + f + 2 h)(c_{\theta_S}^2 + 3) s_{\theta_S}   - 3 e c_{\theta_S}  (c_{\theta_S}^2 - 3)\right] \\
%\ea
%\ee
%\be 
%\ba{rl}
\lambda_{H_2H_2H_2}& = 6v \left(\lambda_5s\omega_3 + \lambda_6 c\omega_3 \right) , \\
\lambda_5 &= 2\left[(b+f+2 h)c_{\theta_S}^2 +6(c+g) s_{\theta_S}^2 +3e c_{\theta_S}   s_{\theta_S}\right]  \\
\lambda_6 &= -4\left[6 a c_{\theta_S}^3+s_{\theta_S}^2 ((b+f+2 h) c_{\theta_S}  +3 e s_{\theta_S} )\right] \\
%\ea
%\ee
%\be 
%\ba{rl}
\lambda_{h_0h_0H_1} &= 2v \left(\lambda_7s\omega_3 + \lambda_8 c\omega_3 \right) , \\ 
\lambda_7 &=\displaystyle\frac{1}{4}\left[  3(b + f + 2 h) (3 c_{\theta_S}  + 1) s_{\theta_S}^2  \right. \\
& \left. + 2(c + g) (c_{\theta_S}  (3 c_{\theta_S}  (9 c_{\theta_S}  + 1) - 5) + 27)  - 
  3e (9 c_{\theta_S} ^2 - 3 c_{\theta_S}  + 5)  s_{\theta_S}\right]  \\
\lambda_8 &= \displaystyle\frac{1}{4}\left[ 54 a s_{\theta_S}^3+ (b+f+2 h)(9 c_{\theta_S} ^2-5) s_{\theta_S} -3e c_{\theta_S}  (9 c_{\theta_S} ^2+5)\right] \\ 
\ea
\ee
\be 
\ba{rl}
\lambda_{h_0h_0H_2} &= 2v \left(\lambda_9s\omega_3 + \lambda_{10} c\omega_3 \right) ,\\
\lambda_9 &=  \displaystyle\frac{1}{2}\left(2 s_{\theta_S}  ( 3 (b (1-c_{\theta_S} )+  (9 c-f+9 g-2 h) c_{\theta_S} -2 c+f-2 g+2 h) c_{\theta_S} +g)\right. \\
& \left.+ 3(b+f+2 h) s_{\theta_S} ^3 +2 c s_{\theta_S} -9e (2 c_{\theta_S} +1)  s_{\theta_S} ^2+3e c_{\theta_S}  (3 c_{\theta_S}  (c_{\theta_S} +1)-1) \right)  \\
\lambda_{10} &= 6(-9 a+b+f+2 h) c_{\theta_S}  s_{\theta_S} ^2 -(b+f+2 h)(3 c_{\theta_S} ^3+c_{\theta_S} ) -3e (9 c_{\theta_S} ^2-1)  s_{\theta_S}  \\
%\ea
%\ee
%\be 
%\ba{rl}
\lambda_{h_0H_1H_1} &= 2v \left(\lambda_{11}s\omega_3 + \lambda_{12} c\omega_3 \right) , \\ 
\lambda_{11} &=\displaystyle\frac{\sqrt{3}}{4}\left(   (b + f + 2 h)(3 c_{\theta_S}  + 5) s_{\theta_S} ^2  + 
  2 (c + g) (9 c_{\theta_S} ^3 + 5 c_{\theta_S} ^2 + c_{\theta_S}  - 27) \right. \\
& \left.  + 3e ((5 - 3 c_{\theta_S} ) c_{\theta_S}  + 1)  s_{\theta_S}  \right)\\
\lambda_{12} &=\displaystyle\frac{\sqrt{3}}{2}\left(18 a s_{\theta_S} ^3+(b+f+2 h)(3 c_{\theta_S} ^2+1) s_{\theta_S}  -9e c_{\theta_S} ^3 +3e c_{\theta_S}   \right) \\
%\ea
%\ee
%\be 
%\ba{rl}
\lambda_{h_0H_1H_2} &= v \left(\lambda_{13}s\omega_3 + \lambda_{14} c\omega_3 \right) , \\
\lambda_{13} &=   \sqrt{3}\left(-2   (  (b(1+ c_{\theta_S}) -9 c c_{\theta_S} -2 c+c_{\theta_S}  f-9 c_{\theta_S}  g+2 (c_{\theta_S} +1) h+f-2 g) c_{\theta_S} +g) s_{\theta_S} \right.\\
&\left. +(b+f+2 h) s_{\theta_S} ^3 -2 c s_{\theta_S} +3e (1-2 c_{\theta_S} )  s_{\theta_S} ^2+3e c_{\theta_S}  ((c_{\theta_S} -1) c_{\theta_S} +1)\right)  \\
\lambda_{14} &= 2\sqrt{3}\left(  (2  (-9 a+b+f+2 h)s_{\theta_S} ^2 - (b+f+2 h)(c_{\theta_S} ^2-1)-9e c_{\theta_S}   s_{\theta_S} )c_{\theta_S}-3 e s_{\theta_S}   \right) \\
\ea
\ee
\be 
\ba{rl}
\lambda_{h_0H_2H_2} &= 2v \left(\lambda_{15}s\omega_3 + \lambda_{16} c\omega_3 \right) , \\ 
\lambda_{15} &=   \sqrt{3}\left( -2  (  (b+f-9 g+2 h)c_{\theta_S}-9 c c_{\theta_S} +c+g)s_{\theta_S} ^2 +(b+f+2 h)(c_{\theta_S} -1) c_{\theta_S} ^2 \right.\\
& \left.+3e c_{\theta_S}  (2 c_{\theta_S} +1)  s_{\theta_S} -3 e s_{\theta_S} ^3\right)  \\
\lambda_{16} &= 2\sqrt{3}s_{\theta_S}\left( 18 a c_{\theta_S} ^2-(b+f+2 h)(2 c_{\theta_S} ^2 -s_{\theta_S} ^2 )-9e c_{\theta_S}   s_{\theta_S}   \right) \\
%\ea
%\ee
%\be 
%\ba{rl}
\lambda_{H_1H_1H_2} &= 2v \left(\lambda_{17}s\omega_3 + \lambda_{18} c\omega_3 \right) , \\ 
\lambda_{17} &= \displaystyle\frac{1}{2}\left(  -2   (  (b (c_{\theta_S} +3)+  (-9 c+f-9 g+2 h)c_{\theta_S} -6 c+3 f-6 g+6 h)c_{\theta_S}-3 g)s_{\theta_S}\right. \\
&\left. +(b+f+2 h)s_{\theta_S} ^3 +6 c s_{\theta_S} +3e (3-2 c_{\theta_S} )  s_{\theta_S} ^2+3e c_{\theta_S}  ((c_{\theta_S} -3) c_{\theta_S} -3)\right)  \\
\lambda_{18} &=  2(-9 a+b+f+2 h) c_{\theta_S}  s_{\theta_S} ^2 -(b+f+2 h)  (c_{\theta_S} ^2+3) c_{\theta_S}-9e (c_{\theta_S} ^2-1)  s_{\theta_S}    \\
%\ea
%\ee
%\be 
%\ba{rl}
\lambda_{H_1H_2H_2} &= 2v \left(\lambda_{19}s\omega_3 + \lambda_{20} c\omega_3 \right) \, , \\ 
\lambda_{19} &=  2 (- (b+f-9 g+2 h)c_{\theta_S} +c (9 c_{\theta_S} +3)+3 g)s_{\theta_S} ^2 \\
& + (b+f+2 h)(c_{\theta_S} +3) c_{\theta_S} ^2  +3e c_{\theta_S}  (2 c_{\theta_S} -3)  s_{\theta_S} -3 e s_{\theta_S} ^3  \\
\lambda_{20} &= 2 \left( 18 a c_{\theta_S} ^2 - (b+f+2 h)(2  c_{\theta_S} ^2- s_{\theta_S} ^2) - 9e c_{\theta_S}   s_{\theta_S} \right) \, .
\ea
\ee

%

%%%%%%%%%%%%%%%%%%%
%	A C K N O W L E D G M E N T S
%%%%%%%%%%%%%%%%%%%
\begin{acknowledgments}
We thank F. Gonz\'alez Canales for useful discussions regarding the quark sector of SM$\otimes S(3)$.This work has been partially supported  by \textit{CONACYT-SNI (M\'exico)}.
\end{acknowledgments}

%%%%%%%%%%%%%%%%%%%
%	BIBLIOGRAPHY
%%%%%%%%%%%%%%%%%%%
%\bibliographystyle{JHEP}

%%%%%%%%%%%%%%%%%%%%%%%%%%%%%%%%%%%%%%%%%%%%%%%%%%

\end{document}